\documentclass{appolb}
\usepackage{epsfig}
\usepackage{wrapfig,rotating}
\usepackage{amssymb}
\usepackage{amsmath}

\newcommand{\PO}{\rm l \! P }

\newcommand{\xpom}{x_{\PO} }

\newcommand{\be}{\begin{equation}}
\newcommand{\ee}{\end{equation}}

\begin{document}
\title{QCD analysis of diffractive phenomena %
\thanks{Presented at PIC, Annecy, France, 26 - 29 june 2007}%
}
\author{Laurent SCHOEFFEL
\address{CEA Saclay/DAPNIA-SPP, 91191 Gif-sur-Yvette, France}
}
\maketitle
\begin{abstract}

The most important results on subnuclear diffractive phenomena 
obtained at HERA and Tevtaron are reviewed and 
new issues in nucleon tomography are discussed.
Some challenges for understanding
diffraction at the LHC, including the discovering of the Higgs boson, are outlined.

\end{abstract}
  
\section{Introduction}

One of the most important experimental results from the DESY electron-proton collider HERA,
working at a center of mass energy of about 300 GeV,
is the observation of a significant fraction, around $15\%$, of 
large rapidity gap events in deep inelastic scattering (DIS). In these events,
the target proton emerges in the final state with a loss of a very small
fraction ($\xpom$) of its energy-momentum \cite{royon}. 

\begin{figure}[hpt]
\begin{center}
\includegraphics[width=0.7\textwidth]{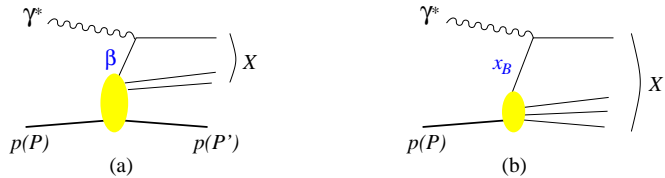}
\caption{Parton model diagrams for deep inelastic diffractive (a) and
  inclusive (b) scattering observed at lepton-proton collider HERA. The variable $\beta$ is the momentum
  fraction of the struck quark with respect to $P-P'$, and the Bjorken variable $x_{Bj}$ its
  momentum fraction with respect to $P$.}
\label{fig1}
\end{center}
\vspace{-0.5cm}
\end{figure}

In Fig. \ref{fig1}(a), we present this event topology,
$\gamma^* p \rightarrow X \ p'$, where the virtual photon $\gamma^*$ probes the proton structure and originates from the
electron. Then, the final hadronic state $X$ and the scattered proton are well separated in
space (or rapidity) and a gap in rapidity can be observed in the event with no particle produced 
between $X$ and the scattered proton. 
In the standard QCD description
of DIS,  such events are not expected in such an abundance since large
gaps are exponentially suppressed due to color strings formed between
the proton remnant and scattered  partons (see Fig. \ref{fig1}(b)). 
The theoretical description of these processes, also called diffractive processes, is a
real challenge since it must combine perturbative QCD effect of hard scattering with
nonperturbative phenomenon of rapidity gap  formation \cite{lolo}. 
The name diffraction in high-energy particle physics originates from the
analogy between optics and nuclear high-energy
scattering. In the Born approximation the equation for hadron-hadron elastic
scattering amplitude can be derived from the scattering of a plane wave
passing through and around an absorbing disk, resulting in an optic-like diffraction
pattern for hadron scattering. 
The quantum numbers of the initial beam particles are conserved during the reaction
and then 
the diffractive system is  well separated in rapidity from the scattered hadron.

\begin{figure}[hpt]
\vspace{-0.5cm}
\begin{center}
\includegraphics[width=1.\textwidth]{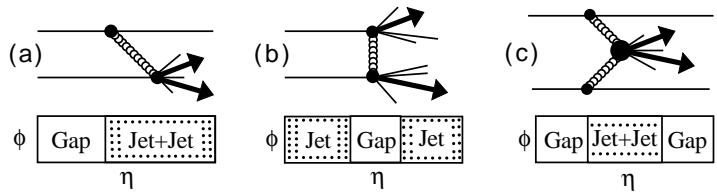}
\vspace*{-1cm}
\caption{ Schematic diagrams of topologies  representative of
hard diffractive processes studied by the proton-antiproton collider Tevatron.}
\label{fig2} 
\end{center}
\vspace{-0.5cm}
\end{figure}

The discovery of large rapidity gap events at HERA has led to
a renaissance of the physics of diffractive scattering in an
entirely new domain, in which the large momentum transfer 
 provides a hard scale. 
This observation has also revived the the
rapidity gap physics with hard triggers, as large-$p_{\perp}$ jets, at the
proton-antiproton collider Tevatron, currently working at a center of mass energy of about 2 TeV (see Fig. \ref{fig2})
\cite{royon,lolo}.

Whether
the existence of such  hard scales makes the diffractive processes
tractable within the perturbative QCD  or not has been a subject of
intense theoretical and experimental research during the past
decade. In the following, we describe the main ideas and results. 
Using the standard vocable, the vaccum/colorless exchange involved in the diffractive interaction
is called Pomeron in this paper.

\section{Basics of diffractive interactions}

\begin{figure}[!]
\begin{center}
\includegraphics[width=10cm,height=8.5cm]{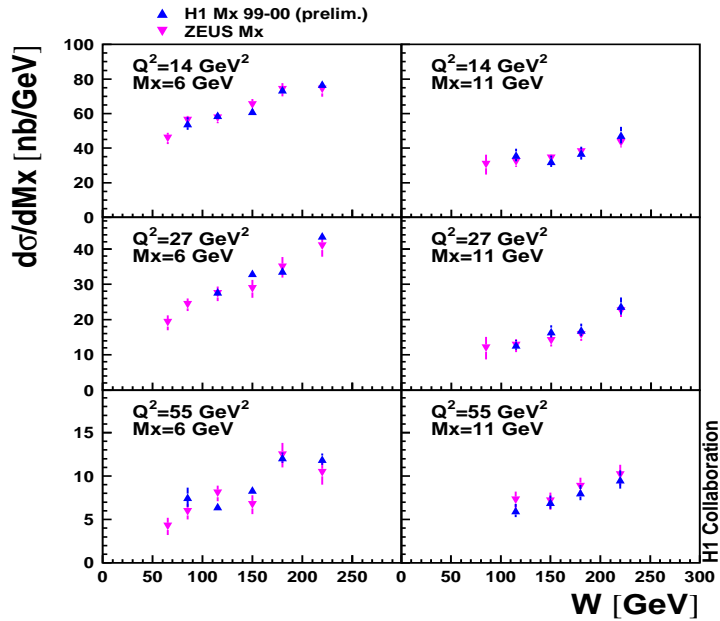}
\caption{The  cross section of the diffractive process $\gamma^* p \rightarrow p' X$, 
differential in the mass of the diffractively produced hadronic system $X$ ($M_X$),
is presented as a function of the centre-of-mass energy of the $\gamma^*p$ system $W$.
Measurements at different values of the virtuality
$Q^2$ of the exchanged photon are displayed.
}
\label{figdata}
\end{center}
\vspace{-0.5cm}
\end{figure}

The inclusive diffractive cross section has been measured at HERA by H1 and
ZEUS experiments over a wide kinematic range,
as illustrated in Fig. \ref{figdata}. We observe that the diffractive cross section
shows a hard dependence in the centre-of-mass energy of the $\gamma^*p$ system $W$.
Namely, we get a behaviour of the form $\sim W^{ 0.6}$  for the diffractive cross section, 
compatible with the dependence expected
for a hard process. This  first observation allows further studies of the diffractive process in the context of
perturbative QCD.

\begin{wrapfigure}{r}{0.3\columnwidth}
\centerline{\includegraphics[width=0.28\columnwidth]{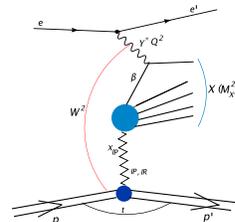}}
\caption{Kinematics.}\label{kin}
\end{wrapfigure}

Several theoretical formulations have been proposed to describe the  diffractive exchange.
The purpose here is  to describe the "blob" displayed in Fig. \ref{fig1}(a) in a quatitative 
way, leading to a proper description of data shown in Fig. \ref{figdata}.
Among the most popular models, the one based on a pointlike structure of
the Pomeron   assumes that the exchanged object, the Pomeron, 
is a colour-singlet quasi-particle whose structure is probed in
the reaction. In this approach, diffractive parton distribution functions (PDFs)
are derived from the diffractive DIS cross sections 
in the same way as standard PDFs are extracted from DIS measurements \cite{lolo}.
It means that a certain flux of Pomeron is emitted off the proton, depending on 
the variable $\xpom$, the fraction of the longitudinal momentum of
the proton lost during the interaction (see Fig. \ref{kin}).
Then, the partonic structure of the Pomeron is probed by 
the diffractive exchange (see Fig. \ref{fig1}(a) and \ref{kin}).  In Fig. \ref{kin}, we illustrate this factorisation property and
remind the notations for the kinematic variables used in this paper, as
the virtuality
$Q^2$ of the exchanged photon,
 the centre-of-mass energy of the $\gamma^*p$ system $W$ and
 $M_X$ the mass of the diffractively produced hadronic system $X$.
It follows that the Bjorken variable $x_{Bj}$ verifies $x_{Bj} \simeq Q^2/W^2$ in the 
low $x_{Bj}$ of the H1/ZEUS measurements ($x_{Bj}<0.01$). Also,  the Lorentz invariant variable $\beta$
defined in Fig. \ref{fig1} is equal to $x_{Bj}/\xpom$ and can be interpreted as
the fraction of longitudinal momentum of the struck parton in the (resolved) Pomeron.

This resolved Pomeron model gives a good description of
HERA data but fails to describe the Tevatron results.
Indeed, some underlying interactions can occur during the proton-antiproton collision,
which break the gap in rapidity produced in the diffractive process \cite{royon}.

\section{Dipole model of diffractive interactions}
\begin{figure}[!]
  \vspace*{-1.5cm}
     \centerline{
         \psfig{figure=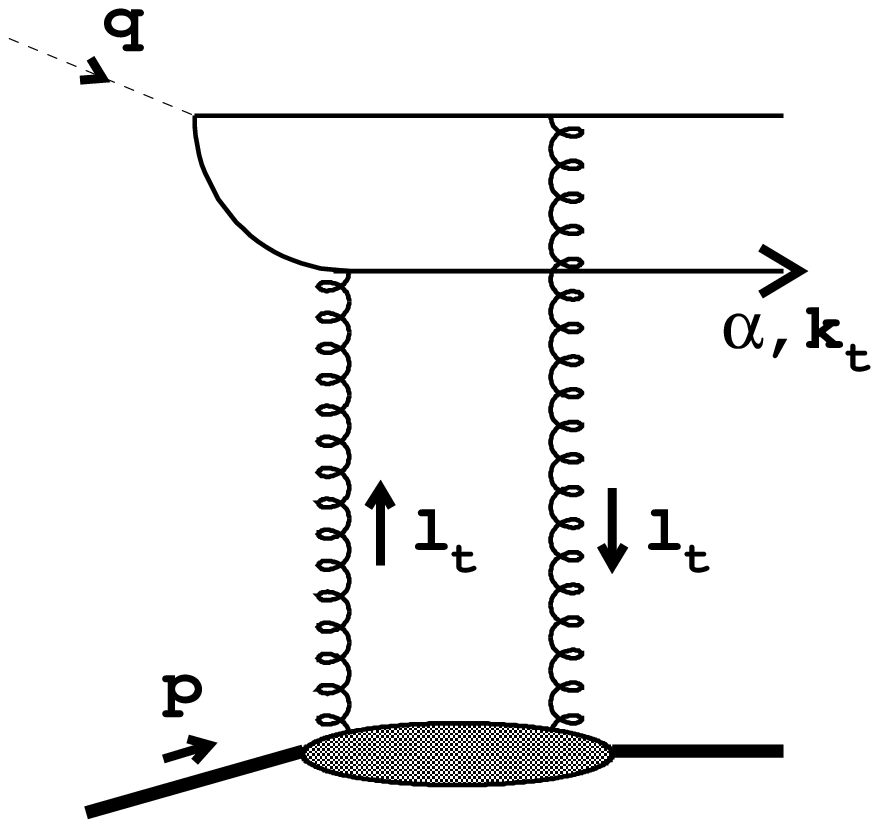,width=5cm}
         \psfig{figure=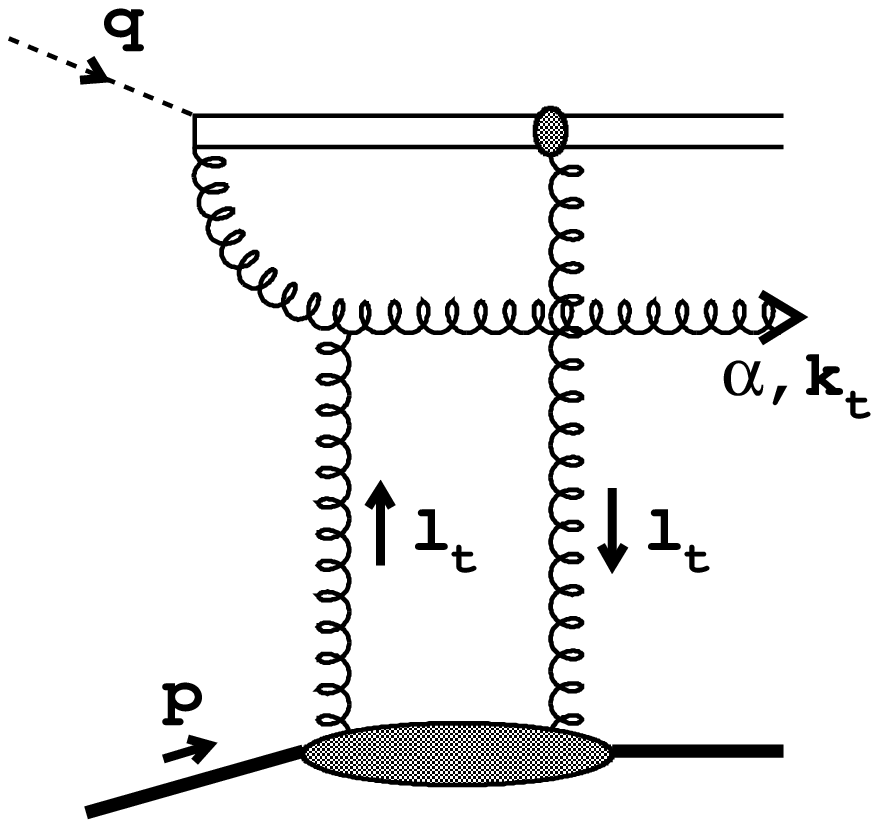,width=5cm}
           }
\vspace*{-0.5cm}
\caption{\it The $q\bar{q}$ and $q\bar{q}g$ components of the diffractive system.}
\label{fig4}
\end{figure}

In the following, we focus our discussion on 
a different approach of diffractive interactions in which the the process is modeled with the 
 exchange of (at least) two gluons projected onto the color singlet state (see Fig. \ref{fig4})
 \cite{dipole1,dipole3}. 
In this model, the reaction follows three different phases displayed in Fig. \ref{fig4} :
(i) the
transition of the virtual photon to the $q\bar{q}$ pair (the color
dipole) at a large distance
$
l \sim {1\over m_N x}
$ of about 10-100 fm for HERA kinematics,
upstream the target, 
(ii) the interaction of the
color dipole with the target nucleon, and (iii) the projection of
the scattered $q\bar{q}$ onto the the diffractive system $X$.

\begin{figure}[!]
\begin{center}
\includegraphics[width=12cm,height=6.5cm]{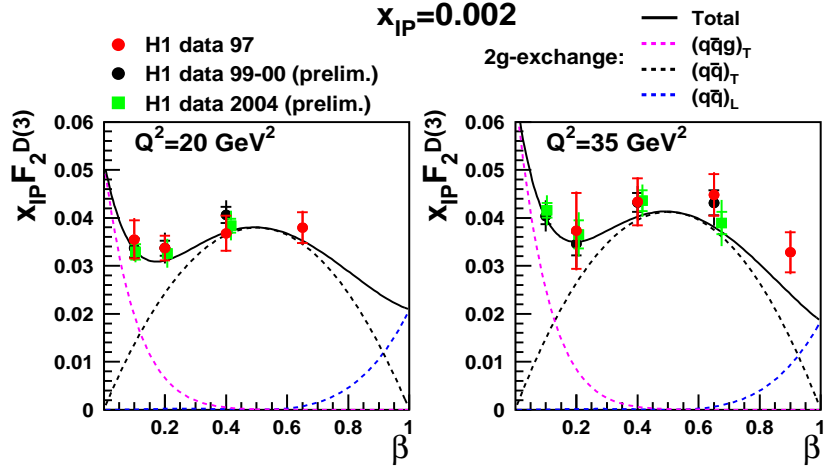}
\caption{ The diffractive structure function
$\xpom F_2^{D(3)}$ is presented  as a function of $\beta$
for two values 
of $Q^2$. The different components of the two-gluon
exchange model are displayed (see text). They add up to give a good description of the data. The structure function 
  $ \xpom F_2^{D(3)}$
is obtained directly from the
measured  diffractive cross section using the relation : 
$
\frac{d^3 \sigma^{ep\rightarrow eXp}}{d\xpom\ dx\ 
dQ^2} \simeq \frac{4\pi\alpha_{em}^2}{xQ^4}
({1-y+\frac{y^2}{2}}) F_2^{D(3)}(\xpom,x,Q^2)
$, where $y$ represents the inelasticity of the reaction.}
\label{figbekw}
\end{center}
\end{figure}
The inclusive diffractive
cross section is then described with
three main contributions \cite{bartels}. The first one describes the diffractive production of a $q \bar{q}$ pair from
a transversely polarised photon, the second one the production of 
a diffractive $q \bar{q} g$ system, and the third one the production of a
$q \bar{q}$ component from a longitudinally polarised photon (see Fig. \ref{fig4}).
In Fig. \ref{figbekw}, we show that this two-gluon exchange model gives a good description
of the diffractive cross section measurements. 

\begin{figure}[!]
   \centering
   \epsfig{file=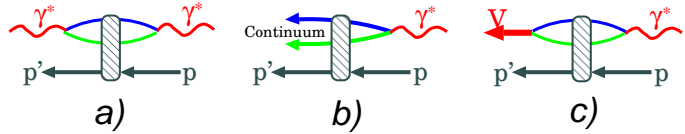,width=90mm}
   \caption{\it    The unified picture of Compton scattering,
diffraction excitation of the photon into hadronic continuum
states and into the diffractive vector meson
\label{figunified}}
\end{figure}

One of the great interest of the two-gluon exchange approach
is that it provides a unified description of different kind of processes measured in $\gamma^* p$ collisions at HERA 
\cite{ivanov} :
inclusive $\gamma^* p \rightarrow X$, diffractive $\gamma^* p \rightarrow X  \ p'$
and (diffractive) exclusive vector mesons (VM) production $\gamma^* p \rightarrow VM \ p'$
(see Fig. \ref{figunified}). In the last case, the step (iii) described above consists
in the recombination of
the scattered pair $q\bar{q}$ onto a real VM (as $J/\Psi$, $\rho^0$, $\phi$,...) or onto 
a real photon for the reaction $\gamma^* p \rightarrow \gamma \ p'$,
which is called deeply virtual Compton scattering (DVCS).

\begin{wrapfigure}{r}{0.5\columnwidth}
\centerline{\includegraphics[width=0.45\columnwidth]{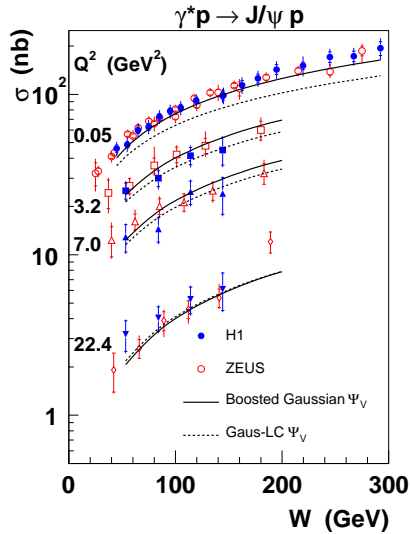}}
\caption{$J/Psi$ exclusive production cross section as a function of $W$ 
  for different $Q^2$ values
  compared to predictions from a dipole model (including
  saturation ans skewing effects). 
  Two different vector meson wave functions are used corresponding to the two curves.}
  \label{fig:crossw}
\end{wrapfigure}

The dipole model then predicts a strong rise of the cross section in $W$, which reflects the rise at small $x_{Bj}$
of the gluon density in the proton.
Indeed, in the two-gluon exchange model, the exclusive VM production cross section can be simply expressed as
proportional to the square of the gluon density. 
At low $x_{Bj}$, the gluon density 
increases rapidly when $x_{Bj}$ decreases and therefore a rapid increase of the cross section
with $W$ is expected and observed (see Fig. \ref{fig:crossw}).
Note that saturation effects,
that are screening the large increase
of the dipole cross section (gluon density) at low $x_{Bj}$,
are taken into account in recent developments of the dipole approach. 
Also, skewing effects are included in models, i.e. 
the difference between the proton momentum fractions carried by the two 
exchanged gluons. 
The gluon density is then replaced by a generalized parton distribution, 
labeled $F_g$ in the following. The VM cross section is then related to the square of $F_g$.
Finally, a reasonable agreement with the data is obtained (see Fig. \ref{fig:crossw}) \cite{ivanov, kowalski}.

\section{Nucleon tomography }
\begin{figure}[!]
\begin{center}
\includegraphics[width=6cm,height=4.cm]{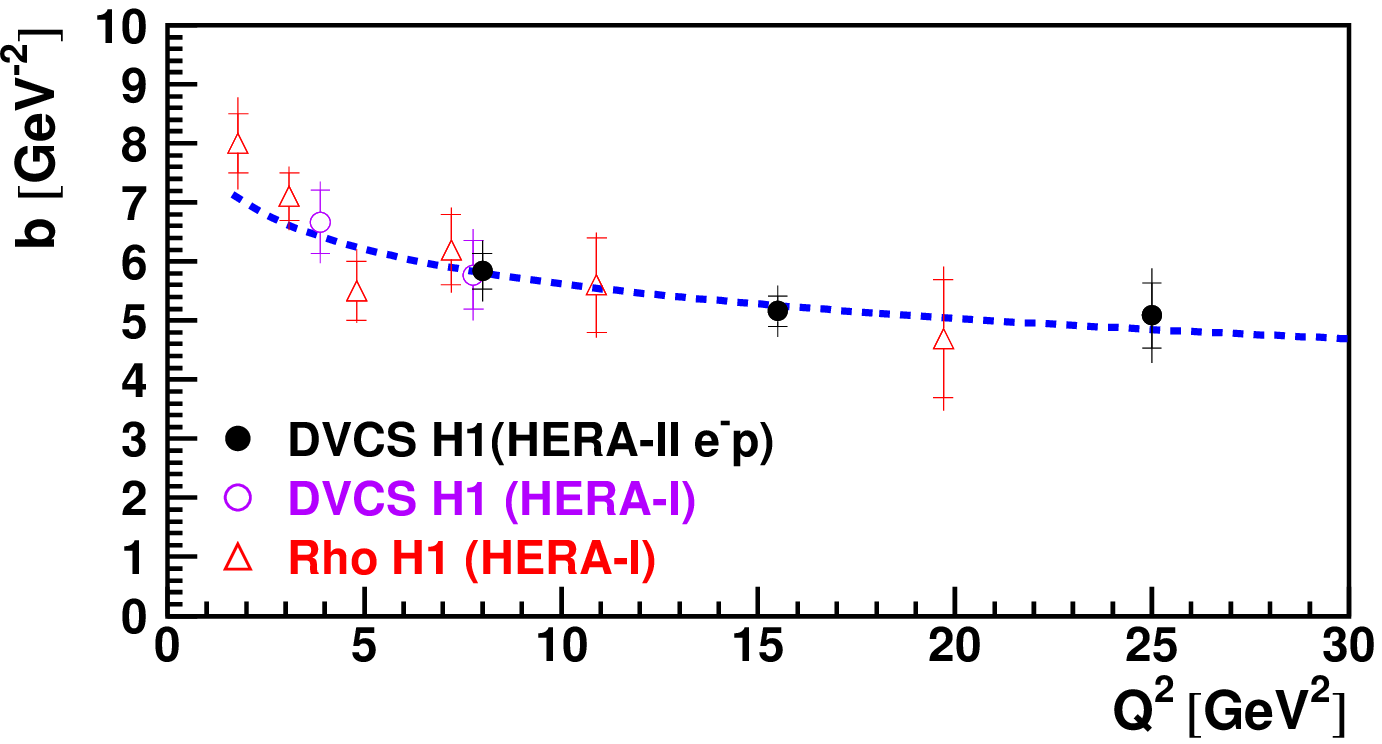}
\includegraphics[width=6cm,height=4.cm]{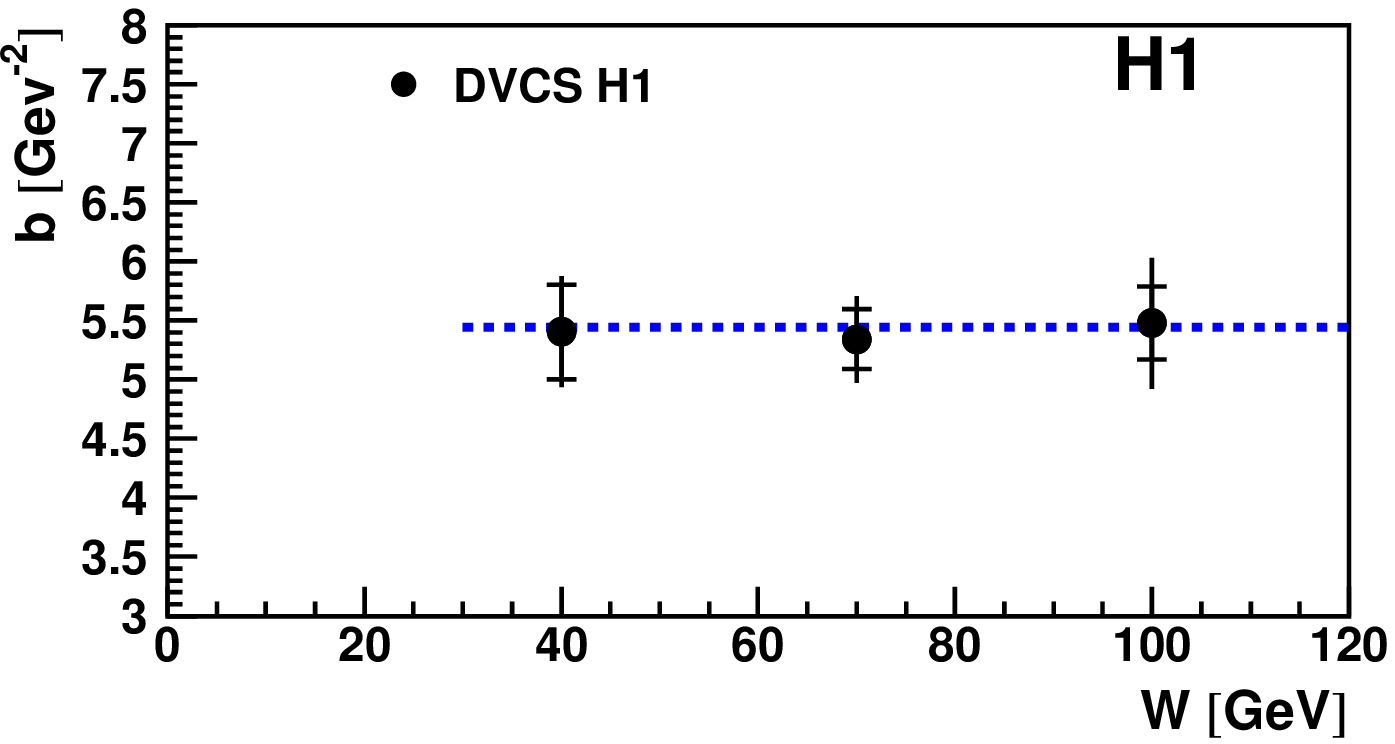}
\caption{ The logarithmic slope of the $t$ dependence
  for DVCS and $\rho$ exclusive production :
  $d\sigma/dt \propto
\exp(-b|t|)$  where $t=(p-p')^2$.
}
\label{bslopes}
\end{center}
\vspace{-0.7cm}
\end{figure}

One of the key measurement in exclusive processes is the slope
defined  by  the exponential fit to the differential cross section:  
$
d\sigma/dt \propto
\exp(-b|t|)
$
at small $t$, where $t=(p-p')^2$ is the square of the momentum transfer at the
proton vertex (see Fig. \ref{bslopes}). 
A Fourier transform from momentum
to impact parameter space readily shows that the $t$-slope $b$ is related to the
typical transverse distance between the colliding objects \cite{buk,diehl}.
At high scale, the $q\bar{q}$ dipole is almost
point-like, and the $t$ dependence of the cross section is given by the transverse extension 
of the gluons (or sea quarks) in the  proton for a given $x_{Bj}$ range.
More precisely, from the  generalised gluon distribution $F_g$ defined in section 3, we can compute
a gluon density which also depends on a spatial degree of freedom, the transverse size (or impact parameter), labeled $R_\perp$,
in the proton. Both functions are related by a Fourier transform 
$$
g (x, R_\perp; Q^2) 
\;\; \equiv \;\; \int \frac{d^2 \Delta_\perp}{(2 \pi)^2}
\; e^{i ({\Delta}_\perp {R_\perp})}
\; F_g (x, t = -{\Delta}_\perp^2; Q^2).
$$
Thus, the transverse extension $\langle r_T^2 \rangle$
 of gluons (or sea quarks) in the proton can be written as
$$
\langle r_T^2 \rangle
\;\; \equiv \;\; \frac{\int d^2 R_\perp \; g(x, R_\perp) \; R_\perp^2}
{\int d^2 R_\perp \; g(x, R_\perp)} 
\;\; = \;\; 4 \; \frac{\partial}{\partial t}
\left[ \frac{F_g (x, t)}{F_g (x, 0)} \right]_{t = 0} = 2 b
$$
where $b$ is the exponential $t$-slope.
Measurements of  $b$
have been performed  for  different channels, as DVCS or $\rho$ production (see Fig. \ref{bslopes}-left-),
which corresponds to $\sqrt{r_T^2} = 0.65 \pm 0.02$~fm at large scale $Q^2$ for $x_{Bj} \simeq 10^{-3}$ \cite{lsdvcs,lsdvcs2}.
This value is smaller that the size of a single proton, and, in contrast to hadron-hadron scattering, it does not expand as energy $W$ increases
(see Fig. \ref{bslopes}-right-).
This result is consistent with perturbative QCD calculations in terms of a radiation cloud of gluons and quarks
emitted around the incoming virtual photon.

\section{Quarks total angluar momenta }

In section 3, we have introduced the generalised parton distributions (GPDs) in the
presence of skewing, difference of momenta between the two exchanged gluons or quarks.
These functions have interesting features : they interpolate between the standard PDFs
and hadronic form factors. Also, they complete the nucleon spin puzzle as they provide a measurement
of the total angular momentum contribution of any parton to the nucleon spin \cite{Ji:1996nm}.
In the DVCS process, the skewing is large. It can be shown that the difference
of momenta between the two exchanged partons reads : $\delta(x) = x_{Bj}/(2-x_{Bj})$.
That's why it is a golden reaction to access GPDs, in particular when
measuring its interference with the non-discernable  (electro-magnetic) Bethe-Heitler  (BH) process. 

\begin{wrapfigure}{r}{0.5\columnwidth}
\centerline{\includegraphics[width=0.48\columnwidth]{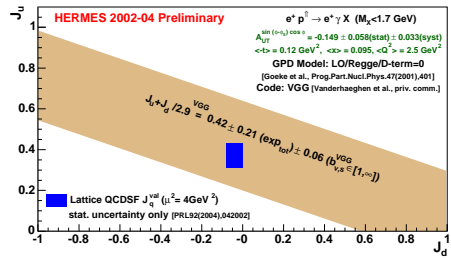}}
\caption{Constraint on $u$-quark total angular momentum $J_u$ vs $d$-quark total angular momentum $J_d$, 
 obtained at HERMES. Also shown is a Lattice result.}\label{hermes}
\end{wrapfigure}
Following this strategy, different asymmetries can be 
extracted \cite{lsdvcs,lsdvcs2,hermes1}, which depend on the helicity/charge of the beam particles
or on the polarisation of the target. These asymmetries are then directly related to the DVCS/BH interference,
hence directly sensitive to GPDs.
The HERMES collaboration, which is a  fixed target
experiment using the 27.6 GeV electron beam of HERA,  has completed recently a measurement
direclty sensitive to the  $u$ and $d$ quarks angular momenta \cite{hermes2}. The result is given in
Fig. \ref{hermes} where for the first time a constraint in the plane $J_u/J_d$ can be derived \cite{hermes2}.

\section{Towards LHC}

In recent years, the production of the Higgs boson in diffractive 
proton-proton collisions at the LHC has drawn more and more attention as a clean channel to 
study the properties of a light Higgs boson or even discover it. This is 
an interesting example of a new  challenge. The idea is to search for exclusive events
at the LHC, as illustrated in Fig. \ref{fig2}(c) \cite{royonlhc,lslhc}.
The full energy available
in the center of mass is then used to produce the heavy object, which can be a dijet system,
a $W$ boson or could be a Higgs boson.
With this topology, the event produced is very clean : 
both protons escape and are detected in forward Roman pot detectors, two large rapidity gaps are created
on both sides and the central production of the  heavy object gives some decay products well
isolated in the detector (see Fig. \ref{fig2}(c)).
A second advantage of such events is  that the resolution on the mass of
the  produced object can be determined with a high resolution from the
measurement of the proton momentum loses ($x_{\PO,1}$ and $x_{\PO,2}$), using the relation $M^2= s x_{\PO,1} x_{\PO,2}$
where $\sqrt{s}$ is the center of mass energy available in the collision.
A potential signal, accessible in a mass distribution, is then not washed out by the lower resolution
when using central detectors, rather than forward Roman pots to measure $x_{\PO,1}$ and $x_{\PO,2}$
\cite{royonlhc,lslhc}.

\section{Conclusions }

We have presented and discussed the most recent results on
 diffraction from the HERA and Tevatron experiments.
With exclusive processes studies, we have shown that a scattering system consisting
of a small size vector particle and the proton has a transverse extension (at high scale) smaller than a single proton 
and does not expand as energy increases.
This result is consistent with perturbative QCD calculations in terms of a radiation cloud of gluons and quarks
emitted around the incoming virtual photon.
Of special interest for future prospects is the exclusive
production  heavy objects (including Higgs boson) at the LHC.


\end{document}